\documentclass{amsart}

\usepackage{graphicx}
\theoremstyle{definition}

\theoremstyle{remark}

\numberwithin{equation}{section}

\begin{document}

\title{Source coding model for repeated snapshot imaging}

\author{Junhui Li$^1$}
\address{$^1$ State Key Laboratory of Advanced Optical Communication Systems and Networks, School of Electronics Engineering and Computer Science, and Center for Quantum Information Technology, Peking University, Beijing 100871, China}

\author{Bin Luo$^2$}
\address{$^2$ State Key Laboratory of Information Photonics and Optical Communications, Beijing University of Posts and Telecommunications, Beijing 100876, China}

\author{Dongyue Yang$^3$}
\address{$^3$ School of Electronic Engineering, Beijing University of Posts and Telecommunications, Beijing 100876, China}

\author{Guohua Wu$^3$}

\author{Longfei Yin$^1$}

\author{Hong Guo$^1$}
\email{hongguo@pku.edu.cn}



\begin{abstract}
Imaging based on successive repeated snapshot measurement is modeled as a source coding process in information theory. The necessary number of measurement to maintain a certain level of error rate is depicted as the rate-distortion function of the source coding. Quantitative formula of the error rate versus measurement number relation is derived, based on the information capacity of imaging system. Second order fluctuation correlation imaging (SFCI) experiment with pseudo-thermal light verifies this formula, which paves the way for introducing information theory into the study of ghost imaging (GI), both conventional and computational. 
\end{abstract}

\maketitle

Repeated snapshot imaging, which replaces a single long time shot with many successive repeated snapshots of short duration, is a major solution to several practical problems, including the insufficient dynamic range of commercial digital cameras, the harsh requirements on stability and noise control of long-exposure measurement, and the high speed imaging (see, e.g., \cite{HDRbook}, pp. 148). It is natural to expect for a more accurate image at the cost of taking more snapshots. The study on how many snapshots one needs to recover an image with certain level of accuracy, i.e., a particular error rate (ER), is crucial for the imaging design, assessment, and optimization. However, few has been reported \cite{arxiv}. 

Ghost imaging (GI), both conventional \cite{Shih95}, and computational \cite{Shapiro08}, is essentially a repeated snapshot imaging. The huge number of measurement, $n$, to maintain image quality has become a major problem preventing GI from practical applications, even if using the compressive sensing (CS) technique to recover the image with fewer snapshots by increasing the calculation cost \cite{Bromberg09}. Unfortunately, though different perspectives of the image quality assessment problem has been investigated \cite{Shapiro09,Lugiato10,SNR11}, study on the ER vs. $n$ relation of GI is quite rare. There are reports on experimental phenomenon and magnitude estimation based on computational complexity e.g., \cite{IEEE08}, yet few trial demonstrates a quantitative formula. On the other hand, information theory has been proven to be a power tool for studies on the quality assessment of both the optical imaging process \cite{SNR2011}, and the CS technique \cite{CS06}, thus provides a promising perspective to study the ER vs. $n$ problem of GI. 

In this Letter, a repeat snapshot imaging process is described by the source coding model in information theory (see, e.g., \cite{Viterbibook}, Chap. 7). The correspondence is built between two relations: the necessary measurement number $n$ versus the image error rate (ER), and the minimum coding rate $R$ under a given distortion $D$, i.e., the rate-distortion function $R \left( D \right)$, which can be derived by the mutual information between the source and the user, i.e., the object and the image in our case. Based on the information capacity of an imaging system \cite{Cox86}, a formula of $R \left( D \right)$ is derived, which results in a quantitative model of the ER vs. $n$ relation. Second order fluctuation correlation imaging (SFCI) experiment, which is a mimic of GI, under pseudo-thermal light illumination, verifies this relation upon different forms of ER. Considering the equivalence of GI and SFCI \cite{SingleArm07,Jane14}, this ER vs. $n$ relation, and more generally, this modeling method, should also apply to GI. 

\begin{figure}[htbp]
\centering
\includegraphics[width=\linewidth]{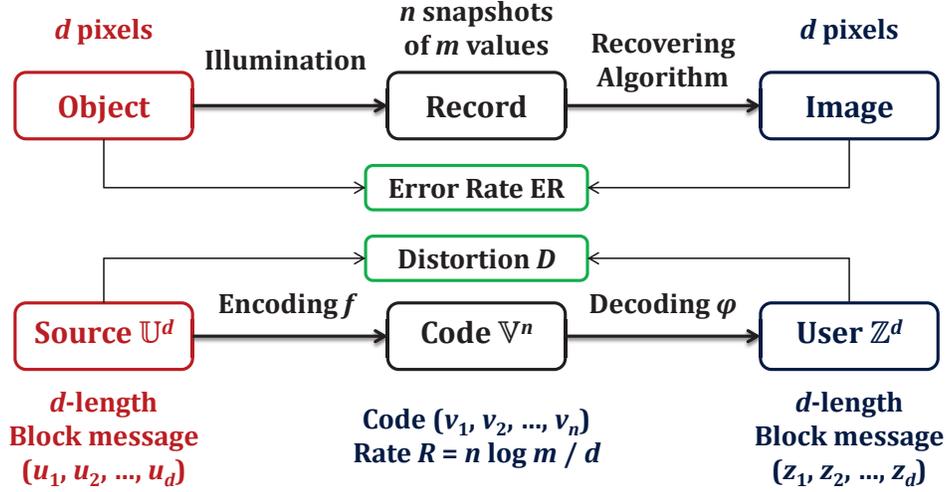}
\caption{Scheme of repeated snapshot imaging and its source coding model. The solution to the source coding problem, i.e., to find the quantitative form of the distortion-rate function $R \left( D \right)$, depicts the ER vs. $n$ relation.}
\label{fig:1}
\end{figure}

The repeated snapshot imaging process and its source coding model is shown in Fig. \ref{fig:1}. The object is illuminated in a certain way to form a record sequence made by $n$ snapshots. The record of each snapshot has $m$ possible outcomes. The image of $d$ pixels is recovered by an algorithm corresponding to the illumination. For simplicity, we assume the object also has $d$ pixels. The quality of the recovered image is assessed by the error rate ER between the image and the object. A source coding model is developed accordingly. A block message of length $d$, $U=\left({u_1},{u_2}, \ldots ,{u_d}\right)$, standing for all the object pixels, experiences an encoding $f$, which describes the illumination process, and turns $U$ in to a code $V=\left({v_1},{v_2}, \ldots ,{v_n}\right)$, representing the record made by $n$ snapshots. The code rate $R = {{n\log m} \mathord{\left/ {\vphantom {{n\ln m} d}} \right. \kern-\nulldelimiterspace} d}$ is proportional to the sampling rate ${n \mathord{\left/ {\vphantom {n d}} \right. \kern-\nulldelimiterspace} d}$ of the imaging process, where $n$ is the measurement number. After a decoding process $\varphi$, which stands for the recovering algorithm, a $d$-length block message $Z=\left({z_1},{z_2}, \ldots ,{z_d}\right)$ is recovered, representing the image. Between the message sent by the source and that received by the user, the coding process $\left( {f,\varphi} \right)$ introduces a distortion $D$, which depicts how much of the received message is wrong, is the counterpart of ER. 

Our goal is to get a quantitative formula of the ER vs. $n$ relation. In the source coding model, it turns into the question that what is the required code rate $R$ if one wants to keep the distortion no more than $D$, i.e., to find a quantitative formula of the rate-distortion function $R \left( D \right)$. This is known as the source coding problem in the information theory (\cite{Viterbibook}, Sec. 7.1). The source coding theorem (\cite{Viterbibook}, Theorem 7.2.1) and the converse source coding theorem (\cite{Viterbibook}, Theorem 7.2.3) indicate that $R \left( D \right)$ equals to the minimum of the mutual information $I\left( {U:Z} \right)$ among all the conditional probability distributions $P\left( {\left. U \right|Z} \right)$ which hold a distortion no more than $D$ (\cite{Viterbibook}, Eq. (7.2.53), please note that though being equivalent, here we use $P\left( {\left. U \right|Z} \right)$ instead of $P\left( {\left. Z \right|U} \right)$). Since $R \left( D \right)$ is a continuous, strictly decreasing function of $D$ (\cite{Viterbibook}, Lemma 7.6.1), the minimum is fulfilled when $D \left( {U, Z} \right) = D$, where the distortion $D\left( { \cdot , \cdot } \right)$ corresponds to the error rate ER between the object and the image in our case. Therefore, the rate-distortion function $R \left( D \right)$ reads,
\begin{eqnarray}
\label{eq:1}
R\left( D \right) &=& \mathop {\min }\limits_{D\left( {U,Z} \right) \le D} I\left( {U:Z} \right) = I\left( {U:Z} \right)\left| {_{D\left( {U,Z} \right) = D}}  \right. \nonumber\\
 &=& H\left( U \right) - H\left( {U\left| Z \right.} \right),
\end{eqnarray}
where the Shannon entropy $H \left( U \right)$ stands for the amount of information of the object, and the conditional entropy $H\left( {U\left| Z \right.} \right)$ depicts how much information of the object is not been revealed by the image. Now the problem is to determine the form of $H \left( U \right)$ and $H\left( {U\left| Z \right.} \right)$.  

Hereafter the information capacity of imaging systems, $N$, is introduced, which is the logarithm of the total number of possible states, $\Omega$ \cite{Cox86},
\begin{equation}
\label{eq:2}
N = \log \Omega = - \sum {\frac{1}{\Omega }\log \frac{1}{\Omega }} ,
\end{equation}
where the second $=$ is based on the ergodicity assumption that each possible state is equally approachable, and indicates that $N$ can be regarded as a Shannon entropy. Meanwhile, for a two dimension image on the $x-y$ plane, the total number of possible states, $\Omega$, reads,
\begin{equation}
\label{eq:3}
\Omega = {\left( {1 + {\rm{SNR}}} \right)^{{\rm{TBP}} \times {\rm{SBP}}}},
\end{equation}
where ${\rm{TBP}} = 1 + 2T{B_T}$ and ${\rm{SBP}} = \left( {1 + 2{L_x}{B_x}} \right)\left( {1 + 2{L_y}{B_y}} \right)$ are the temporal and spatial bandwidth product of the captured image, respectively (see, e.g., \cite{GoodmanBook}, pp. 27. $T$ is the measurement time. $B_T$/$B_x,B_y$ are the temporal/spatial bandwidth. $L_x$ and $L_y$ are the width and height of the image. $\rm{TBP}$ and $\rm{SBP}$ stands for the number of sample points in temporal and spatial domain necessary to specify completely by the imaging system). $\rm{SNR}$ is the signal-to-noise ratio at each pixel, i.e., how many states the imaging system can distinguish at one sample point.   

The image is considered to be perfectly recovered only when it is exactly the same as the object, i.e., $ H\left( {U\left| Z \right.} \right) = 0$. Otherwise error exists. For a practical imaging system, the error comes from two ways: the lack of knowledge caused by insufficient sampling, and the contribution of the inevitable intrinsic noise, the amplitude of which can be seen as a constant if the source is stationary. The error rate ER depicts how much of the image differs from the object, or equivalently, the proportion of number of different states to the total number of states of the image. After taking away the noise part, it relates directly to the percentage of unrevealed information. Thus as an analogy to Eq. (\ref{eq:2}), the conditional entropy can be expressed as
\begin{equation}
\label{eq:4}
H\left( {U\left| Z \right.} \right) = \log \left[ {\left( {{\rm{ER}} - {k_0}} \right){\Omega _Z}} \right],
\end{equation}
where constant $k_0$ stands for the contribution of noise to ER. Substituting Eq. (\ref{eq:3}), Eq. (\ref{eq:4}) and the expression of the code rate (noticing that there may be a difference of the logarithm base, since the entropy takes base $e$, the expression of $R$ takes the form $R = {{{k_1}n} \mathord{\left/ {\vphantom {{{k_1}n} d}} \right. \kern-\nulldelimiterspace} d}$ with constant $k_1$) into Eq. (\ref{eq:1}), we finally get the ER vs. $n$ formula
\begin{equation}
\label{eq:5}
{\rm{ER}} = {k_0} + \frac{{{\Omega _Z}}}{{{\Omega _U}}}\exp \left( { - {k_1}\frac{n}{d}} \right),
\end{equation}
where $\Omega_U$ and $\Omega_Z$ are the information capacity of the object and the image, respectively. This formula predicts a negative-exponential behavior of the ER vs. $n$ relation, converging to $k_0$ when $n \to \infty$. 

\begin{figure}[htbp]
\centering
\includegraphics[width=\linewidth]{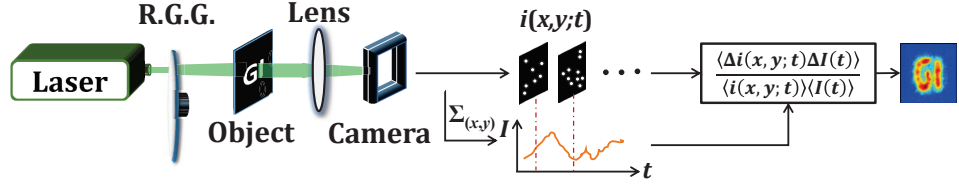}
\caption{The pseudo-thermal light SFCI experiment setup. The image is recovered by Eq. (\ref{eq:6}), and different types of ER are calculated by Eq. (\ref{eq:7}). }
\label{fig:2}
\end{figure}

Pseudo-thermal light SFCI experiment is implemented to verify Eq. (\ref{eq:5}). As shown in Figure \ref{fig:2}, light from a $532~\rm{nm}$ semiconductor laser (Ceo DPSSL-532U, with power less than $32~\rm{mW}$ and linewidth less than $0.1~\rm{MHz}$) passing through a $4~\rm{rpm}$ rotating ground glass plate (Edmund 100 mm diameter $220$ grit) makes a pseudo thermal source with about $30~\rm{ms}$ coherence time (which is smaller than $40~\rm{ms}$ -- the time separation between neighboring measurements, to make sure that the snapshots are independent). This pseudo-thermal light illuminates a static `GI' pattern object of $3~\rm{mm}$ square size (which is much bigger than the coherence length of the light source on the object plane -- $0.02~\rm{mm}$). A imaging lens with $100~\rm{mm}$ focus length forms an image on a commercial 8-bit CMOS camera (Thorlabs DCC 3240C, minimum exposure $0.009~\rm{ms}$), illuminates about $100 \times 100$ pixels. Repeated snapshots under different exposure time and laser power are recorded. The SFCI image is calculated from
\begin{equation}
Z\left( {x,y} \right) \propto \frac{{{{\left\langle {\left[ {i\left( {x,y;t} \right) - {{\left\langle {i\left( {x,y;t} \right)} \right\rangle }_t}} \right] \times \left[ {I\left( t \right) - {{\left\langle {I\left( t \right)} \right\rangle }_t}} \right]} \right\rangle }_t}}}{{{{\left\langle {i\left( {x,y;t} \right)} \right\rangle }_t}{{\left\langle {I\left( t \right)} \right\rangle }_t}}},
\label{eq:6}
\end{equation}
where $i \left({x,y;t} \right)$ and $I \left( t \right) = \sum\nolimits_{x,y} {i\left( {x,y;t} \right)} $ are the spatial record and the `bucket' value, the total count of the snapshot captured at time $t$, respectively. The summation and average denoted by subscript $t$ is over the $N$ snapshots captured. The object $U \left( {x,y} \right)$ is made by taking away the ground glass to capture the image directly illuminated by the laser, then being adjusted into the same size of $Z \left({x,y} \right)$, and being aligned with the center of $Z \left( x,y \right)$. $U \left( {x,y} \right)$ and $Z \left( {x,y} \right)$ are normalized by their total counts, respectively. Two different forms of ER are applied: the mean square error (MSE), and the binary error rate (BER): 
\begin{eqnarray}
\label{eq:7}
{\rm{MSE}} &=& {\left\{ {{{\sum\nolimits_{x,y} {{{\left[ {U\left( {x,y} \right) - Z\left( {x,y} \right)} \right]}^2}} } \mathord{\left/
 {\vphantom {{\sum\nolimits_{x,y} {{{\left[ {U\left( {x,y} \right) - Z\left( {x,y} \right)} \right]}^2}} } {100 \times 100}}} \right.
 \kern-\nulldelimiterspace} \left( {100 \times 100} \right) }} \right\}^{{1 \mathord{\left/
 {\vphantom {1 2}} \right.
 \kern-\nulldelimiterspace} 2}}},\nonumber\\
{\rm{BER}} &=& {{\sum\nolimits_{x,y} {\delta \left( {x,y} \right)} } \mathord{\left/
 {\vphantom {{\sum\nolimits_{x,y} {\delta \left( {x,y} \right)} } {\left( {100 \times 100} \right)}}} \right.
 \kern-\nulldelimiterspace} {\left( {100 \times 100} \right)}},
\end{eqnarray}
where the BER is calculated after $U \left( {x,y} \right)$ and $Z \left( {x,y} \right)$ being changed into binary-value form, $U_B \left( {x,y} \right)$ and $Z_B \left( {x,y} \right)$, respectively, with the average count as the threshold. $\delta \left( {x,y} \right)$ equals to unity if $U_B \left( {x,y} \right) \ne Z_B \left( {x,y} \right),$ being zero otherwise. 
\begin{figure}[htbp]
\centering
\includegraphics[width=\linewidth]{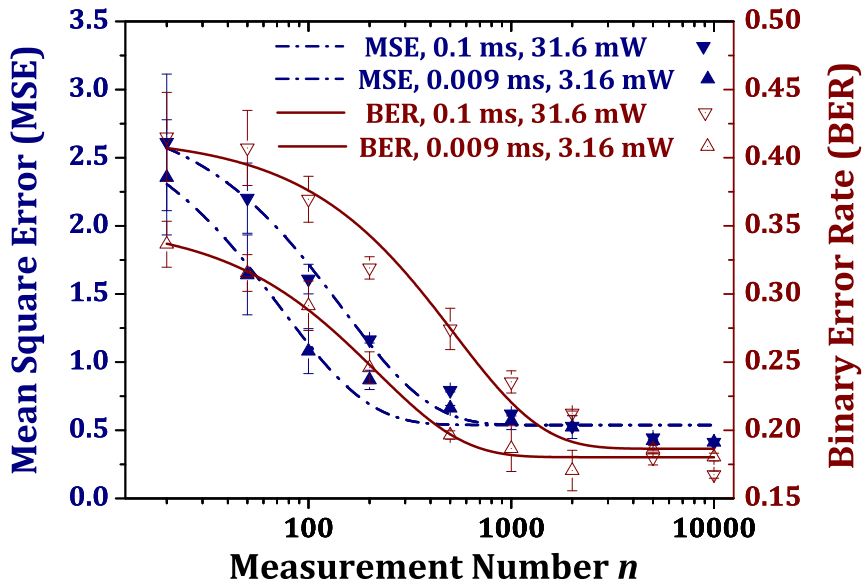}
\caption{Measure ER vs. $n$ relation. The points stand for the measured data, and the lines are fitting curves by Eq. (\ref{eq:7}). Notice that the curves of the same ER type converge to almost the same value when $n \to \infty$.}
\label{fig:3}
\end{figure}

Measured ER upon varying measurement number $n$ are shown as the scatters in Figure \ref{fig:3}. While data under other settings are also collected but not shown, only the two extreme condition with respect to two parameters, i.e., light power and exposure time, are given in Figure \ref{fig:3}. They all fit Eq. (\ref{eq:5}) well, as shown in Table \ref{tab:1}, that each has a close-to-unity adjusted $R$ squared coefficient (see, e.g., \cite{R2}). The range of exposure time is more than $10$, and the large light power is about $10$ times of the small one. Therefore, as for the average number of photons registered by the camera during each snapshot, Eq. (\ref{eq:5}) validates for the range of at least two magnitudes, for both MSE and BER. At the same time, the converge limit of ER when $n \to \infty$, $k_0$, is an invariant, and verifies our assumption in deriving Eq. (\ref{eq:4}). This is reasonable, since $k_0$ stands for the contribution of intrinsic noise to ER. 
\begin{table}[htbp]
\centering
\caption{\bf Fitting Results for ER vs. $n$ by Eq. (\ref{eq:5})}
\begin{tabular}{cccc}
\hline
Type & Power/mW & Adjusted $R^2$ & $k_0$ \\
\hline
MSE & $3.16$ & $0.96613$ & $0.53901 \pm 0.05289$\\
MSE & $31.6$ & $0.98067$ & $0.53759 \pm 0.05330$\\
BER & $3.16$ & $0.99332$ & $0.18648 \pm 0.01077$\\
BER & $31.6$ & $0.96518$ & $0.18027 \pm 0.00264$\\
\hline
\end{tabular}
  \label{tab:1}
\end{table}

\begin{figure}[htbp]
\centering
\includegraphics[width=\linewidth]{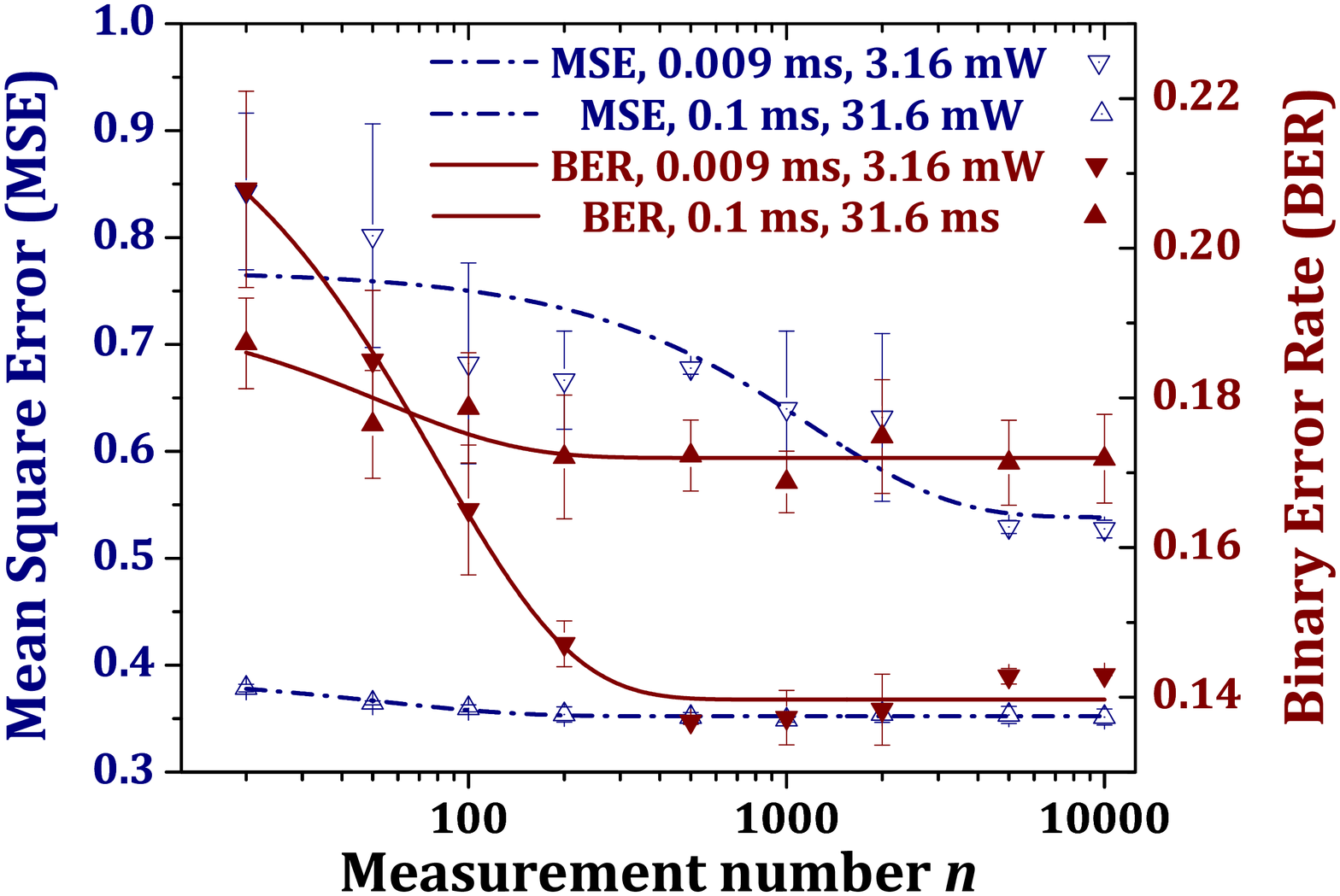}
\caption{Measure ER vs. $n$ relation of DI. The points stand for the measured data, and the lines are fitting curves by Eq. (\ref{eq:7}). }
\label{fig:4}
\end{figure}

Similar experiment is repeated for direct imaging (DI), which recovers the image $Z \left( {x,y} \right)$ directly by the average over $n$ snapshots. The measured ER vs. $n$ relation is shown in Fig. \ref{fig:4}, and the fitting result of Eq. (\ref{eq:5}) is shown in Table \ref{tab:2}. 

We want to mention that all the above fittings by Eq. (\ref{eq:5}) aims only at the negative-exponential behavior, i.e., the fitting function is ${\rm{ER}} = a + b\exp \left( { - cn} \right)$, with all the three constants $a$, $b$, and $c$ adjustable. We only check one of the parameters, $k_0$, in this Letter. Investigation on other parameters of Eq. (\ref{eq:5}) is meaningful. 

At the same time, provided that Eq. (\ref{eq:4}) holds, the negative-exponential behavior does not rely on the particular form of the information capacity $N$, i.e., Eq. (\ref{eq:3}), except that $N$ has the logarithm form like Eq. (\ref{eq:2}). This observation suggests that there is other possible approaches to depict the negative-exponential behavior of the ER vs. $n$ relation. The source coding theorem (\cite{Viterbibook}, Theorem 7.2.1) gives a tight upper bound of the distortion which also demonstrates a similar negative-exponential form. We guess it is the random illumination induced by the pseudo-thermal light, which corresponds to a random encoding $f$ in the source coding model, that makes the ER vs. $n$ formula in our case get that upper bound. But we have not found a solid proof. Meanwhile, an earlier paper by us develops another model which fits the experiment well \cite{arxiv}. That model, which predicts a slightly different behavior, does not involve the source coding theorem at all, but relies exclusively on the particular form of Eq. (\ref{eq:3}). So the question that what is the true origination and deciding factor of the converging behavior of ER, has not reached a solid conclusion yet, since both the form of information capacity (as \cite{arxiv} shows), and the particular method to conduct repeated snapshot imaging, which is essentially a source coding process as shown in this Letter, can lead to this behavior. Meanwhile, there may be other approaches to achieve this negative-exponential behavior. Further investigation is needed to clarify this question. 
\begin{table}[htbp]
\centering
\caption{\bf DI Fitting Assessments for ER vs. $n$ by Eq. (\ref{eq:5})}
\begin{tabular}{cccc}
\hline
Exposure/ms & Power/mW & Type & Adjusted $R^2$ \\
\hline
$0.009$ & $3.16$ & MSE & $0.70469$ \\
$0.009$ & $3.16$ & BER & $0.94946$ \\
$0.1$ & $31.6$ & MSE & $0.98959$ \\
$0.1$ & $31.6$ & BER & $0.75061$ \\
\hline
\end{tabular}
  \label{tab:2}
\end{table}

In conclusion, we built a connection between the source coding system, and the repeated snapshot imaging process. With the help of the theorems on rate-distortion function, and the information capacity of imaging systems, a quantitative formula of the image error rate versus measurement number relation is derived, aiming at predicting the image quality with particular number of snapshots. Under a random illumination induced by a pseudo-thermal light, the formula fits the experiment of both the second order fluctuation correlation imaging and direct average of multiple snapshots well. This formula, and this model in general, should also apply to ghost imaging. 

\subsection*{Funding}
National Science Fund for Distinguished Young Scholars of China (61225003); Natural Science Foundation of China (61471051, 61401036, 61531003, 61571018); Postdoctoral Science Foundation of China (2015M580008); the 863 Program; PhD Students' Overseas Research Program of Peking University, China. 

\bibliographystyle{amsplain}

\end{document}